\documentclass[%
 reprint,
 amsmath,amssymb,
 aps, 
 prd,
]{revtex4-2}

\usepackage{graphicx}
\usepackage{dcolumn}
\usepackage{bm}

\usepackage{xspace}
\usepackage{hyperref}
\usepackage{lineno}
\usepackage{multirow}
\usepackage{xcolor}
\usepackage{booktabs}

\newcommand{\pt}{\ensuremath{p_{\rm T}}\xspace}

\newcommand{\py}{PYTHIA\xspace}

\newcommand{\pPb}{p--Pb\xspace}
\newcommand{\pp}{pp\xspace}

\newcommand*{\Nchj}{\ensuremath{N_\mathrm{j,ch}}\xspace}

\newcommand*{\pT}{\ensuremath{p_\mathrm{T}}\xspace}

\newcommand*{\kT}{\ensuremath{k_\mathrm{T}}\xspace}
\newcommand*{\pTjet}{\ensuremath{p_\mathrm{T}^{\mathrm{jet}}}\xspace}
\newcommand*{\GeV}{\ensuremath{\mathrm{GeV}}\xspace}

\newcommand*{\GeVc}{\ensuremath{\mathrm{GeV}/c}\xspace}

\newcommand*{\LcToDz}{\ensuremath{{\Lambda_{\rm c}^+}/{\rm D^0}}\xspace}
\newcommand*{\LToKzs}{\ensuremath{{\Lambda^0}/{\rm K^0_s}}\xspace}
\newcommand*{\XiToKzs}{\ensuremath{{\Xi^\pm}/{\rm K^0_s}}\xspace}
\newcommand*{\OmToKzs}{\ensuremath{{\Omega^\pm}/{\rm K^0_s}}\xspace}
\newcommand*{\Kzs}{\ensuremath{\rm K_{\rm S}^0}\xspace}
\newcommand*{\pTopi}{\ensuremath{{({\rm p+ \bar{p}})}/{\rm (\pi^{+}+\pi^{-})}}\xspace}
\newcommand*{\jT}{\ensuremath{j_\mathrm{T}}\xspace}
\newcommand*{\etas}{\ensuremath{\eta^\ast}\xspace}
\newcommand*{\phis}{\ensuremath{\varphi^\ast}\xspace}
\newcommand*{\vecps}{\ensuremath{\mathbf{p^\ast}}\xspace}

\begin{document}


\title{Thermodynamical string fragmentation and string density effects in jets}

\author{Róbert Vértesi}
\affiliation{%
    HUN-REN Wigner Research Centre for Physics\\DF13841
    Konkoly-Thege Miklós út 29-33, 1121 Budapest, Hungary
}%
\author{Antonio Ortiz}%
 \email{antonio.ortiz@nucleares.unam.mx}
\affiliation{%
 Instituto de Ciencias Nucleares, UNAM\\
 Circuito exterior s/n Ciudad Universitaria, 04510 Mexico City, Mexico
}%

\date{\today}

\begin{abstract}
It has been proposed to search for thermal and collective properties arising from parton-fragmentation processes by examining high jet charged-constituent multiplicities ($N_{\rm j,ch}$) in proton-proton (pp) collisions. This proposal was initially tested using the PYTHIA~8 event generator with the Monash tune, which incorporates multiparton interactions (MPI) and the MPI-based colour reconnection (CR) model. These studies did not reveal any conclusive evidence for the presence of radial flow.
In this paper, we expand upon the proposed Monte Carlo study by eliminating selection biases associated with triggering on charged particle multiplicities. Furthermore, MPI are disabled to focus exclusively on jet fragments. We analyse pp collisions at $\sqrt{s}$=13 TeV simulated with PYTHIA~8, exploring different implementations of the generator: thermodynamical string fragmentation and the standard Lund fragmentation model. The impact of colour-string juctions  was also explored. Surprisingly, the thermodynamical string fragmentation together with the close-packing of strings mechanism predicts a hint of baryon enhancement in jets. Additionally, the light-flavor baryon-to-meson ratios as a function of $j_{\rm T}$ exhibit similarities across all PYTHIA~8 implementations, and hint at radial flow-like effects. 
In contrast, the ratio of heavy-flavor hadrons (\LcToDz) at low \jT as a function of $N_{\rm j,ch}$ shows a trend similar to that observed as a function of charged-particle multiplicity in minimum-bias data, suggesting that colour-string junctions may play a crucial role in heavy baryon production in jets. The same mechanism also predicts a \jT-integrated \LcToDz ratio that increases with increasing $N_{\rm j,ch}$. Interestingly, for the lowest $N_{\rm j,ch}$ value, such a ratio is consistent with the e$^{+}$e$^{-}$ limit, suggesting that jet multiplicity is a potential way to explore more dilute systems covering the multiplicity gap between  e$^{+}$e$^{-}$ and pp collisions.
\end{abstract}

\maketitle


\section{Introduction}\label{sec:1}

The heavy-ion collider experiments at the Relativistic Heavy Ion Collider (RHIC) at BNL and the Large Hadron Collider (LHC) at CERN aim to study dense QCD systems, where a new form of matter, the strongly-interacting quark--gluon plasma (QGP) is created. One major discovery of the LHC era is the observation of QGP-like effects in small-collision systems (\pp and \pPb collisions). The origin of the new phenomena has remained an open question for the past decade. In heavy-ion collisions, signatures such as collectivity and strangeness enhancement have been attributed to the formation of QGP~\cite{Busza:2018rrf,Bala:2016hlf,ALICE:2022wpn}. On the one hand, IP-Glasma calculations indicate that the energy density, averaged over the transverse area, can reach around 70\,GeV/fm$^{3}$ in \pp collisions for charged particle multiplicity densities above 100~\cite{ALICE:2020fuk}. Lattice QCD calculations suggest that this condition is sufficient for QGP formation in \pp collisions. However, no signatures of jet quenching have been observed so far~\cite{ALICE:2022qxg,ALICE:2023plt}. Moreover, long-range angular correlations have been observed even in very dilute systems like low-multiplicity \pp collisions~\cite{ALICE:2023ulm}, photonuclear ultra-peripheral Pb--Pb collisions~\cite{ATLAS:2021jhn}, and probably in jets~\cite{CMS:2023iam}. Although hydrodynamics has been applied to explain the \pp and \pPb data, some theoretical works suggest that it should not be applicable in small-collisions systems~\cite{Werthmann:2023dvl}. This opens the door to alternative approaches, such as colour strings, which have successfully described some aspects of \pp and \pPb data~\cite{dEnterria:2010xip, Bierlich:2014xba,Bierlich:2018xfw,Ortiz:2024ndh}.

The exploration of diluted QCD systems has continued over the last years. Recently, in Ref.~\cite{Baty:2021ugw} authors postulate that a strongly-interacting QGP-like system can originate from a fast parton (quark or gluon) as it fragments and propagates through the QCD vacuum. Due to the intrinsic QCD coupling strength, the colour fields of the fast parton in the vacuum can produce a large number of secondary partons. These secondary partons can then interact and develop a collective expansion, which is transverse to the trajectory of the parent parton, and extends over a finite space-time volume. The idea is somewhat similar to a Monte Carlo (MC) study that showed that \py produces radial flow-like effects in dense partonic systems originating from multiparton interactions (MPI) due to string interactions via colour reconnection (CR)~\cite{OrtizVelasquez:2013ofg}. Since, in the standard colour reconnection schemes implemented in \py, CR effects are suppressed in jets, no radial flow-like effects and strangeness enhancement were reported in Ref.~\cite{Baty:2021ugw}. 

Inspired by Ref.~\cite{Baty:2021ugw}, the CMS collaboration reported a new measurement from pp collision data with high jet charged-constituents multiplicities (\Nchj)~\cite{CMS:2023iam}. They studied two-particle correlations among the charged particle jet constituents as functions of the azimuthal angle and pseudorapidity separations $(\Delta\varphi^{*},\Delta\eta^{*})$ in the jet coordinate frame, where $\eta^{*}$ and $\varphi^{*}$ are defined relative to the direction of the jet axis. The azimuthal correlation was analysed as a function of \Nchj. For low \Nchj values, the long-range elliptic anisotropic harmonic, $v_{2}^{*}$, decreases with \Nchj, a characteristic reproduced by models like PYTHIA~8. However, an intriguing rising trend for $v_{2}^{*}$ is observed at $\Nchj > 80$, hinting at a possible onset of collective behaviour, which is not reproduced by the MC generators that were tested. The observation is interesting and deserves further study, particularly for high-\Nchj values. 

This paper reports on the search for baryon enhancement and radial flow-like effects in jets. A crucial difference from Ref.~\cite{Baty:2021ugw} is that the bias due to the multiplicity selection based on charged particles is removed by reporting the hyperon-to-${\rm K}_{\rm S}^{0}$ ratios instead of the hyperon-to charged-pion ratios. In addition, the baryon-to-meson ratio in the charm sector is also investigated. For both studies, various models are tested, including the standard Lund string fragmentation model with different colour-reconnection implementations, as well as the thermodynamical string fragmentation model~\cite{Fischer:2016zzs}. The current work is relevant for understanding QGP-like effects observed in small systems, and may also help resolve a long-standing puzzle, that in elementary ${\rm e^{+}e^{-}}$ collisions the yields of various hadron species can be well described by a thermal statistical model~\cite{Becattini:1995hr}.

The paper is organised as follows. The \py MC event generator and the different models used in this work are introduced in section~\ref{sec:2}. Section~\ref{sec:3} presents the methodology, as well as comparisons between models and data. Section~\ref{sec:4} presents the results and discussion. Finally, results are summarised in section~\ref{sec:5}.

\section{\label{sec:2}PYTHIA models}

The results presented in this paper are obtained from \py simulations~\cite{Sjostrand:2014zea} (version 8.309), which incorporate the Lund string fragmentation model~\cite{Andersson:1983ia}. The Monash tune has been obtained from fits to early LHC data; therefore, it successfully describes several observables at LHC energies. The model incorporates MPI and CR, which are crucial for reproducing the multiplicity distributions as well as the correlation between the average transverse momentum (\pt) and the charged-particle multiplicity~\cite{Skands:2014pea}. Since the discovery of radial flow-like effects due to the CR mechanism, more advanced CR models have been built~\cite{Christiansen:2015yqa}. The most common CR models use the leading colour (LC) approximation to calculate the colour flow on an event-by-event basis; the partonic final states result from colour-connecting a quark to another single and unique parton in the event. In the model, gluons carry both a colour and an anticolour charge, and therefore they are connected to two other partons. The LC approximation is used in the MPI-based colour reconnection model implemented in the Monash tune. In this scheme, each MPI-scattering system is viewed as separate and distinct from all other systems in colour space. The colour flow of two such systems can be fused, and if so, the partons of the lower-\pt system are added to the strings defined by the higher-\pt system to minimise the total string length. In this paper we also consider colour reconnection modes beyond leading colour approximation (CR-BLC). The QCD-scheme CR, which considers the minimisation of the string length and the colour rules from QCD, constructs all pairs of dipoles that are allowed to reconnect by QCD-colour rules and switches if the new pair has a lower string length. Junctions can also be directly produced from three, and in some special cases, four dipoles. Here, the multiplet structure of SU(3) is combined with a minimisation of the string potential energy, to decide between which partons strings should form, allowing also for ``baryonic'' configurations. One would expect that MPI increases the number of possible subleading connections.~\cite{Bierlich:2016vgw}. It is worth mentioning that in the traditional approach, also implemented in the MPI-based CR scheme, CR effects are negligible in jets and are enhanced in active MPI environments.
Although the QCD-based CR model implemented in PYTHIA reproduces basic observables well, further tunes have been developed considering different time-dilation constraints~\cite{Christiansen:2015yqa}. The CR-BLC mode 2, which requires strict causality between dipoles, is proven to reproduce the features seen in the heavy-flavor baryonic sector~\cite{ALICE:2021npz,ALICE:2023sgl,CMS:2019uws}.

The models described above are based on the standard Lund string fragmentation, where a tunneling mechanism for string breakups results in a Gaussian suppression of the production of heavier quarks and diquarks, with a further suppression based on the hadronic spin state. In the thermodynamical string fragmentation~\cite{Fischer:2016zzs}, the Gaussian suppression in mass and \pt is replaced by an exponential suppression.  The experimental motivation comes from fixed-target and ISR data, which show that the inclusive \pt distributions are well described by an exponential function, ${\rm exp}(-B p_{\rm T})$ with $B \approx 6$\,GeV$^{-1}$. The model exhibits reasonable agreement with LHC data for observables such as the \pt distributions of unidentified charged hadrons, the average transverse momentum as a function of the hadron mass, and hyperon-to-pion ratios as a function of multiplicity. The set of parameters for thermodynamical-string fragmentation includes the modification of string tension due to the local string density~\cite{Fischer:2016zzs}. This mechanism is named close packing. In minimum-bias pp collisions an effective string density is introduced for low \pt, while jet fragmentation occurs outside the close-packing string region and is left unaffected. This mechanism changes the flavour composition at high multiplicities. In this paper, we explore the potential impact of these new implementations in high-string density environments such as high jet-charged-constituent multiplicities.

The relevant parameters are summarized in Table~\ref{tab:pythia}. In the following, the model named as {\bf Monash CR-QCD} is the Monash tune with the QCD-based CR parameters, which is essentially Monash tune including colour-string junctions. Variations of Monash tune including thermodynamical-string fragmentation with MPI-based and QCD-based CR models are termed {\bf Thermodyn.} and {\bf Thermodyn.~CR-QCD} models, respectively. Finally, the model {\bf CR-BLC mode2} is a tuned version of PYTHIA~8 including colour-string junctions.

\section{\label{sec:3}Analysis method}

\begin{figure*}
	\centering 
	\includegraphics[width=1.0\textwidth]{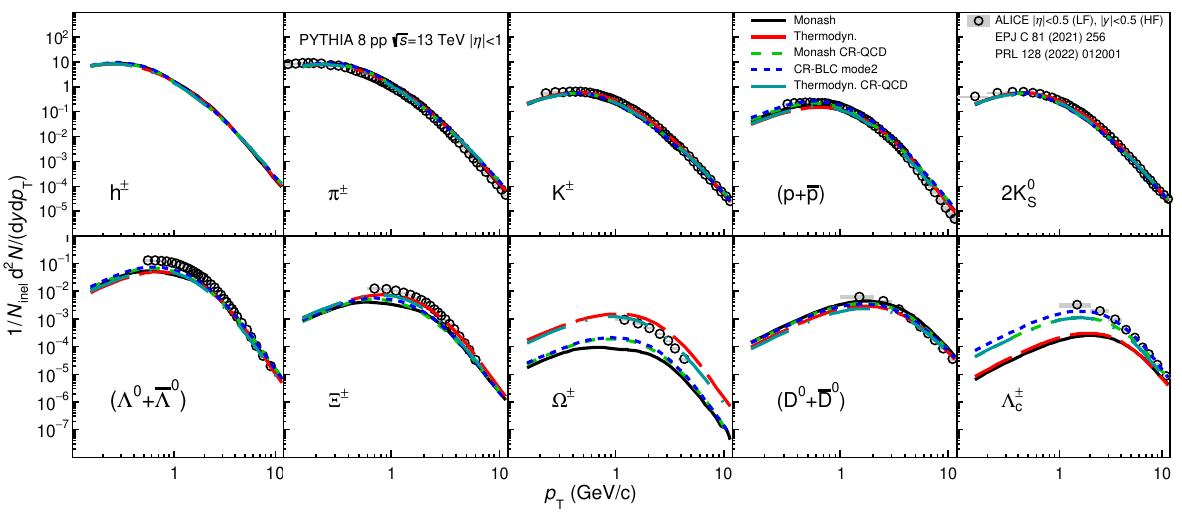}	
	\caption{Midrapidity transverse momentum spectra of identified hadrons in pp collisions at $\sqrt{s}=13$\,TeV. Data are compared with different model predictions.} 
	\label{fig0}%
\end{figure*}

Before discussing the observables in the jet frame, Fig.~\ref{fig0} shows the \pt spectra of various particle species, comparing the model predictions described above with the minimum-bias pp ALICE data~\cite{ALICE:2020jsh,ALICE:2021rzj}. Note that there are certain differences between simulations and data in handling charm hadrons: in the current study, the feed-down components into charm hadrons have not been excluded, and a pseudorapidity cut $|\eta|<1$ was used instead of a rapidity cut $|y|<0.5$ on the charm hadrons, $|\eta|<0.5$ on their decay products as considered in Ref.~\cite{ALICE:2021rzj}.
Overall, all models yield similar results for unidentified charged particles, charged pions, kaons and (anti)protons. The spectral shapes are well reproduced by models both at low and high \pt. For heavier hadrons, the models deviate from each other. The $\Lambda$-\pt spectrum is better described by the Monash~CR-QCD and CR-BLC~mode2 models suggesting the important role of colour-string junctions. The thermodynamical-string fragmentation models predict a slightly lower yield, probably because the model has not been tuned to data. However, the $\Xi^{\rm \pm}$ and $\Omega^{\rm \pm}$ baryon-\pt spectra are best described by the simulations, which consider thermodynamical-string fragmentation. Although the best agreement is observed for $\Omega^{\rm \pm}$, the other models significantly underpredict   particle yield within the measured \pt range (1--4\,GeV/$c$). It is worth noting that the Thermodyn. model matches the data better than the Thermodyn.~CR-QCD model, which overpredicts the yield. For the heavy-flavour sector (${\rm D}^{0}$ meson and $\Lambda_{\rm c}^{\rm \pm}$), the best description is provided by models incorporating colour-string junctions (Monash~CR-QCD, CR-BLC~mode2 and Thermodyn.~CR-QCD), which promote the baryon formation particularly in the charm sector.

For jet observables, the simulations were performed with the HardQCD setting and a phase-space cut of $\hat{p}_{\rm T} > 550\ \GeVc$. Jets were clustered using the anti-\kT algorithm~\cite{Cacciari:2008gp} with the energy recombination scheme and a resolution parameter $R=0.8$. Jets with charged-jet transverse momenta $550 < \pTjet < 1000\ \GeVc$ were used. The mean $\hat{p}_{\rm T}$ is about 744 \GeVc. To correct for the bias introduced by the phase-space cut, we applied a \pTjet-dependent correction, which is in the order of 15\% at $\pTjet=550\ \GeV$ and vanishes towards higher momenta. We made sure through a cross-check with $\hat{p}_{\rm T} > 250\ \GeVc$ that the phase-space cut has negligible effect on our observables. Also note that we applied a selection on \pTjet corresponding to charged jets. Applying full jet transverse-momentum selection similarly to Ref.~\cite{CMS:2023iam} results in a momentum shift, which affects the reach in \Nchj but the baryon-to-meson ratios are largely unaffected. Since we focus on jet observables, we disabled multiparton interactions to avoid background from the underlying event unrelated to the jets. In the analysis, we use the jet frame~\cite{Baty:2021ugw} to define the kinematics of jet constituents as $\vecps = (\jT, \etas, \phis)$, where \jT is the momentum component perpendicular to the jet axis, and \etas-\phis were already defined.

\section{\label{sec:4}Results and discussion}

\begin{figure*}
	\centering 
	\includegraphics[width=0.45\textwidth]{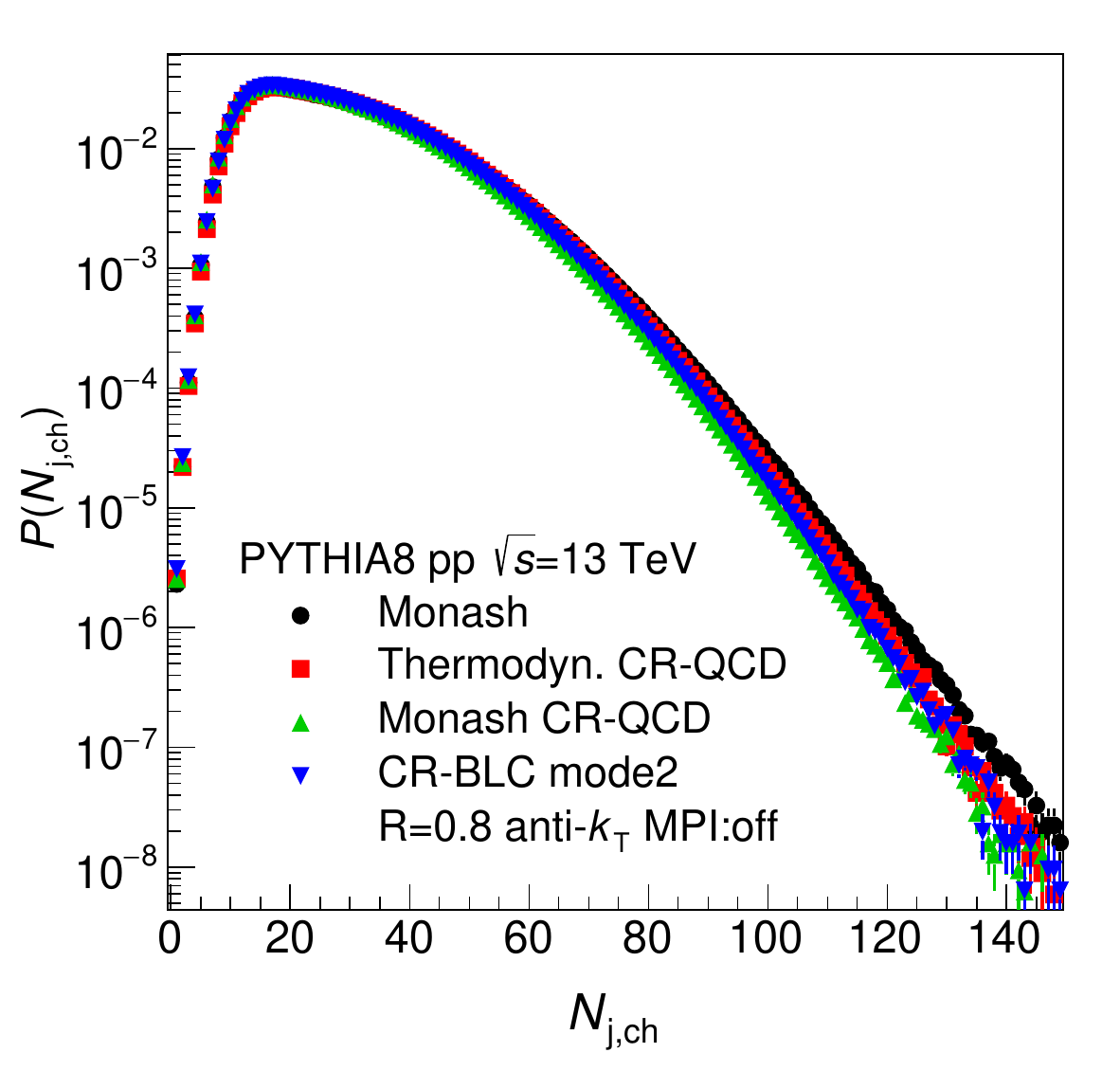}
    \includegraphics[width=0.45\textwidth]{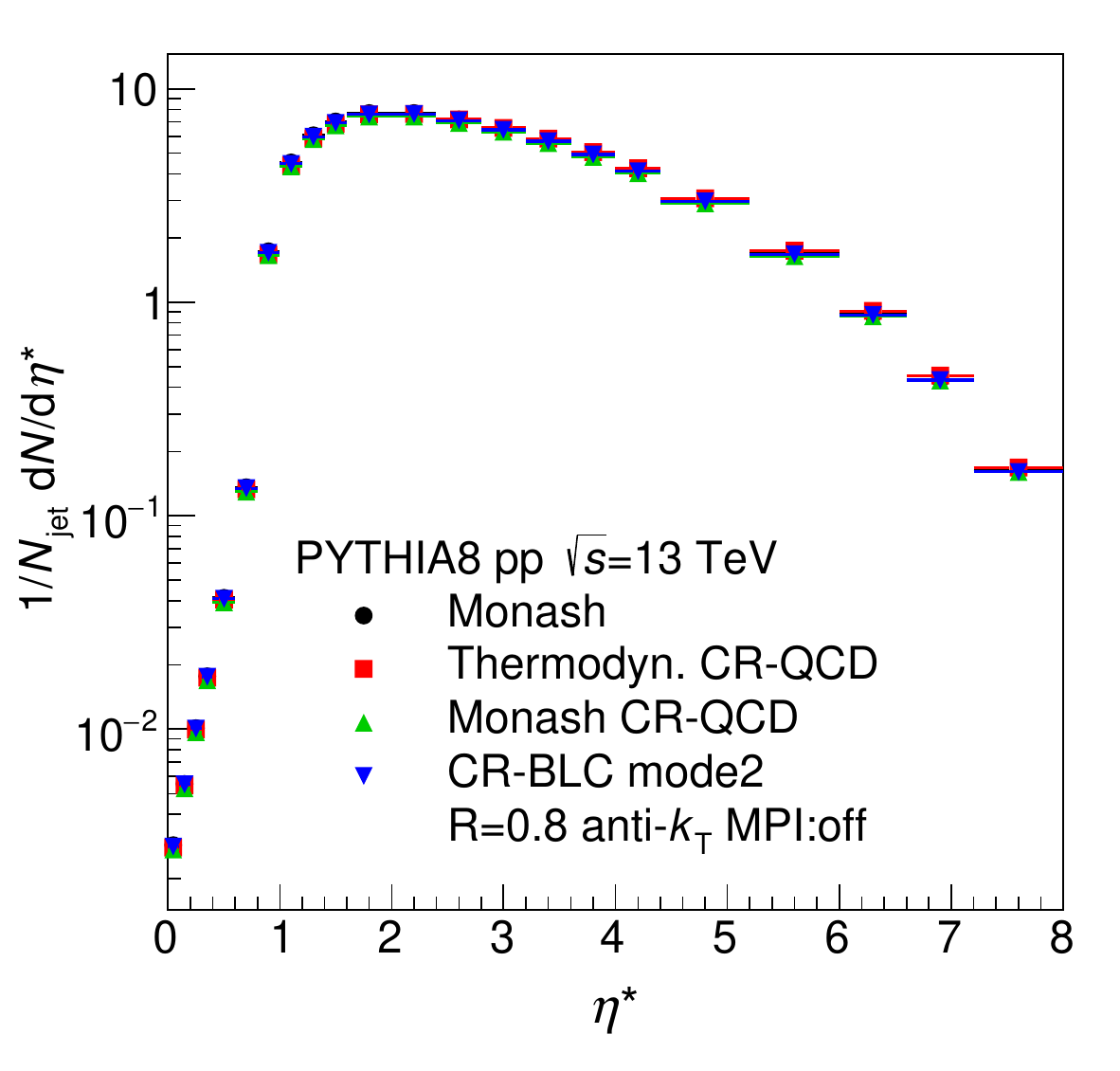}	
	\caption{Jet charged-constituent multiplicity (left) and \etas (right) distributions for different model settings. Results for pp collisions at $\sqrt{s}=13$\,TeV are displayed. MPI is off in all cases.} 
	\label{fig1_2}%
\end{figure*}
Figure~\ref{fig1_2} (left) shows the \Nchj distributions for various model settings. The jet multiplicities show slight variations depending on the colour-reconnection scheme and also differ slightly in the case of thermodynamical string fragmentation. In particular, the high-\Nchj tail is slightly reduced in the Monash~CR-QCD, CR-BLC mode2, and Thermodyn.~CR-QCD models compared to Monash. The \etas distributions, shown in the right panel, appear to be independent of the model settings used. In the jet coordinate system, low-\etas values correspond to charged particles that are separated from the jet axis by large angles, whereas the high-\etas values are associated with particles at small angles relative to the jet direction. All curves exhibit a steep rise up to $\etas \approx 0.86$, which is related to the choice of the jet resolution parameter ($R=0.8$).

\begin{figure*}
	\centering 
	\includegraphics[width=1.0\textwidth]{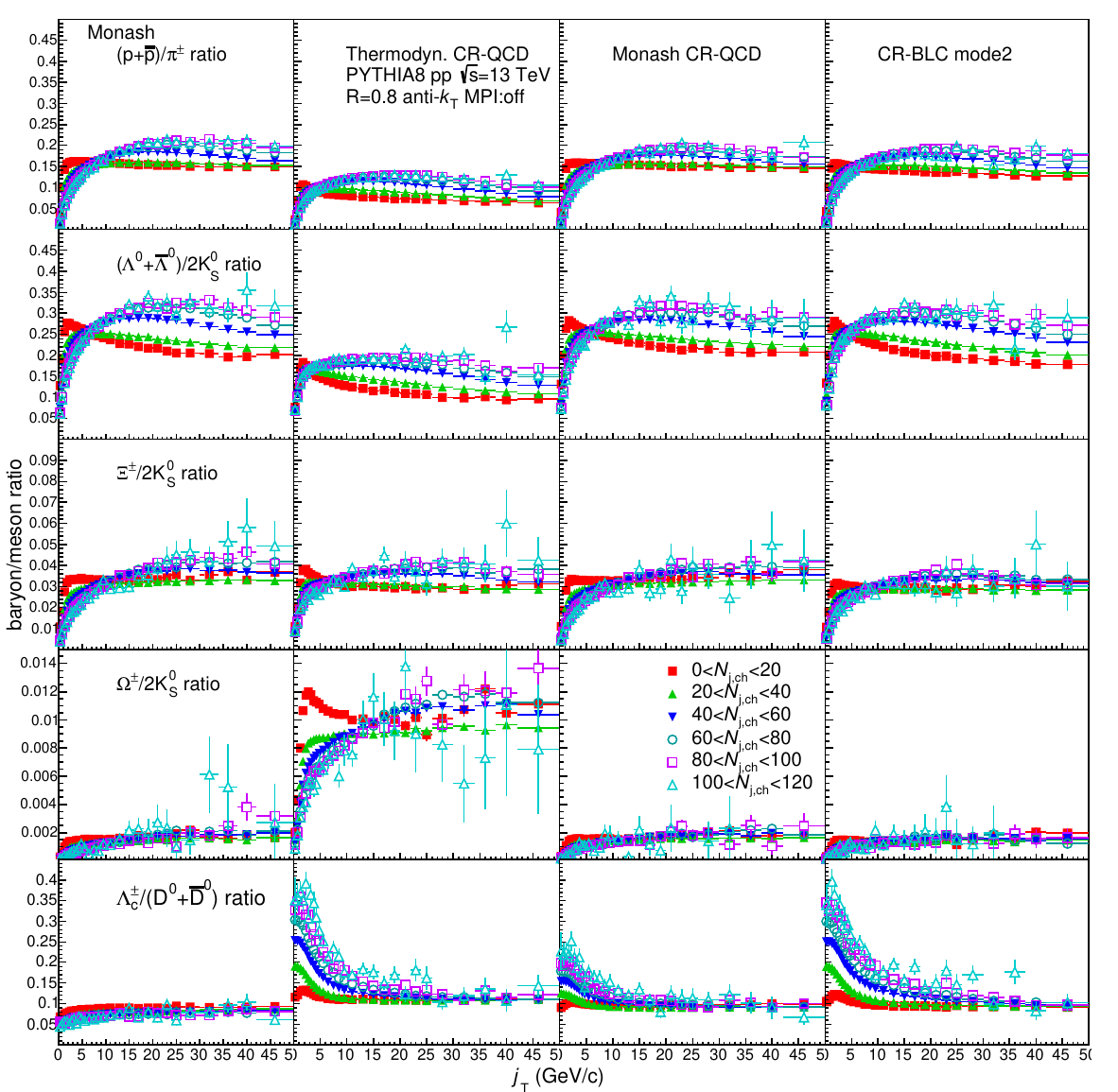}	
	\caption{Baryon-to-meson ratios as a function of \jT for different jet charged-constituent multiplicity intervals. Results are shown for pp collisions at $\sqrt{s}=13$\,TeV, simulated with different \py settings. MPI is off in all cases.} 
	\label{fig3}%
\end{figure*}
To reveal the dynamics of various constituents within a jet, their production can be analysed using kinematic variables in the co-moving frame. Figure~\ref{fig3} shows different baryon-to-meson ratios as a function of \jT, for different models, in several \Nchj classes. 

Before discussing the particle ratios, it is important to note that the models predict larger emission angles and \jT values for high jet charged-constituent multiplicities compared to the inclusive and low-\Nchj event classes. Conversely, lower emission angles (and lower \jT values) are expected for low \Nchj pp collisions.  
This effect is due to a correlation between jet-charged-constituent multiplicity and the flavour of its initiating parton, with high-\Nchj values predominantly associated to gluon-initiated jets and low-\Nchj values to quark-initiated jets~\cite{Baty:2021ugw}.

In the light-flavour sector, all models (Monash, Monash CR-QCD, CR-BLC mode2 and Thermodyn.~CR-QCD) exhibit some common features in baryon-to-meson ratios. For $j_{\rm T}>5$\,GeV/$c$, the proton-to-pion and the $\Lambda^0$-to-\Kzs ratios are grouped into low-\Nchj and high-\Nchj ranges, with the ratios for $0<\Nchj<20$ and $20<\Nchj<40$ being lower than those for the higher \Nchj intervals: 40-60, 60-80, 80-100, 100-120. This grouping may reflect the flavour of the initiating parton. Looking at the details, the ratios are slightly multiplicity dependent, as they increase with increasing \Nchj. On the other hand, for lower \jT values ($\jT<5$\,GeV/$c$) the multiplicity dependence is stronger: the ratios are depleted from the low- to high-\Nchj classes. For all \Nchj classes, the ratios show a bump structure which shifts towards higher \jT values with increasing \Nchj. The effect is similar to that observed in \pt in the standard multiplicity analysis, where it is attributed to radial flow-like effects~\cite{ALICE:2018pal}. All these features, except for the multiplicity grouping, are also present in the $\Xi^\pm$-to-\Kzs particle ratio. The presence of the bump is not obvious in the case of the multistrange $\Omega^\pm$-to-\Kzs ratio, except in the Thermodyn.~CR-QCD model, where a strong enhancement of $\Omega^\pm$ is observed compared to other models. However, this enhancement is less pronounced in high-\Nchj pp collisions at low-\jT, and only gets stronger with \Nchj towards mid- and high-\jT values.

For the heavy-flavour particle ratio, $\Lambda_{\rm c}/{\rm D}^{0}$, no enhancement is seen with the Monash model at low \jT. However, Monash~CR-QCD shows a clear \LcToDz enhancement at lower \jT values, which is more pronounced with increasing \Nchj. This phenomenon is even stronger for the CR-BLC mode2 and Thermodyn.~CR-QCD models. Keeping in mind that string junctions have an impact on the baryon-to-meson ratio at low \pt~\cite{ALICE:2017thy,CMS:2019uws,ALICE:2020wfu,LHCb:2018weo}, the observed patterns in jets could be due to colour-string junctions. One should remember that the \LcToDz ratios as a function of \pt are well described by PYTHIA~8 including colour-string junctions. 
A similar feature to the \Nchj-dependent excess at low \jT values has also been seen in CR-BLC simulations toward higher \pT values, where the \LcToDz ratios were found to depend on the jet activity~\cite{Varga:2021jzb}.
Regarding the CR-BLC~mode2 and Thermodyn.~CR-QCD models, it is also interesting to note that for the highest \Nchj class, the maximum value for \LcToDz is approximately 0.35, similar to the value reported for the lowest inclusive multiplicity class in Ref.~\cite{ALICE:2021npz} suggesting that intra-jet multiplicities is a promising venue to explore the multiplicity gap between e$^+$e$^-$ to pp collisions. This conclusion is further supported by the fact that for the lowest \Nchj class the ratios are nearly flat and consistent with the corresponding ratios measured in ${\rm e}^{+}{\rm e}^{-}$ collisions at LEP, which was found to be $0.113 \pm 0.013{\rm (stat)} \pm 0.006{\rm (syst)}$~\cite{Gladilin:2014tba}. Note that in the low-\Nchj limit, one would expect the jet sample to be dominated by quark-initiated jets. 
It is also important to note that while there are more recent models for colour reconnection with colour junctions~\cite{Bierlich:2023okq,Altmann:2024odn}, there is no significant difference of the $\Lambda_{\rm c}/{\rm D}^{0}$ predictions provided by CR-BLC and these new models. Therefore, we stick to CR-BLC, which is widely used by the community.

\begin{figure*}
	\centering 
	\includegraphics[width=0.5\textwidth]{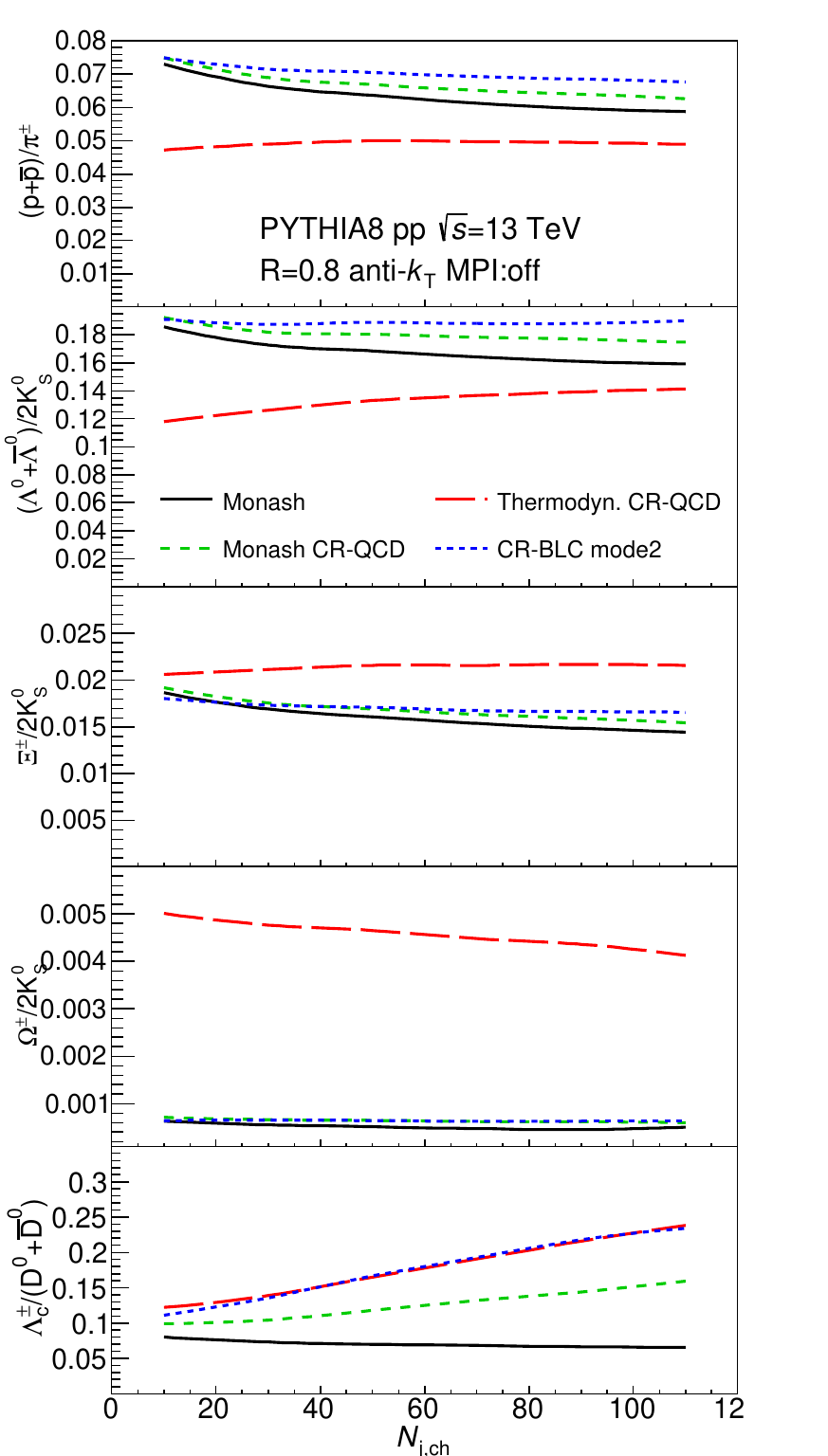}%
	\includegraphics[width=0.5\textwidth]{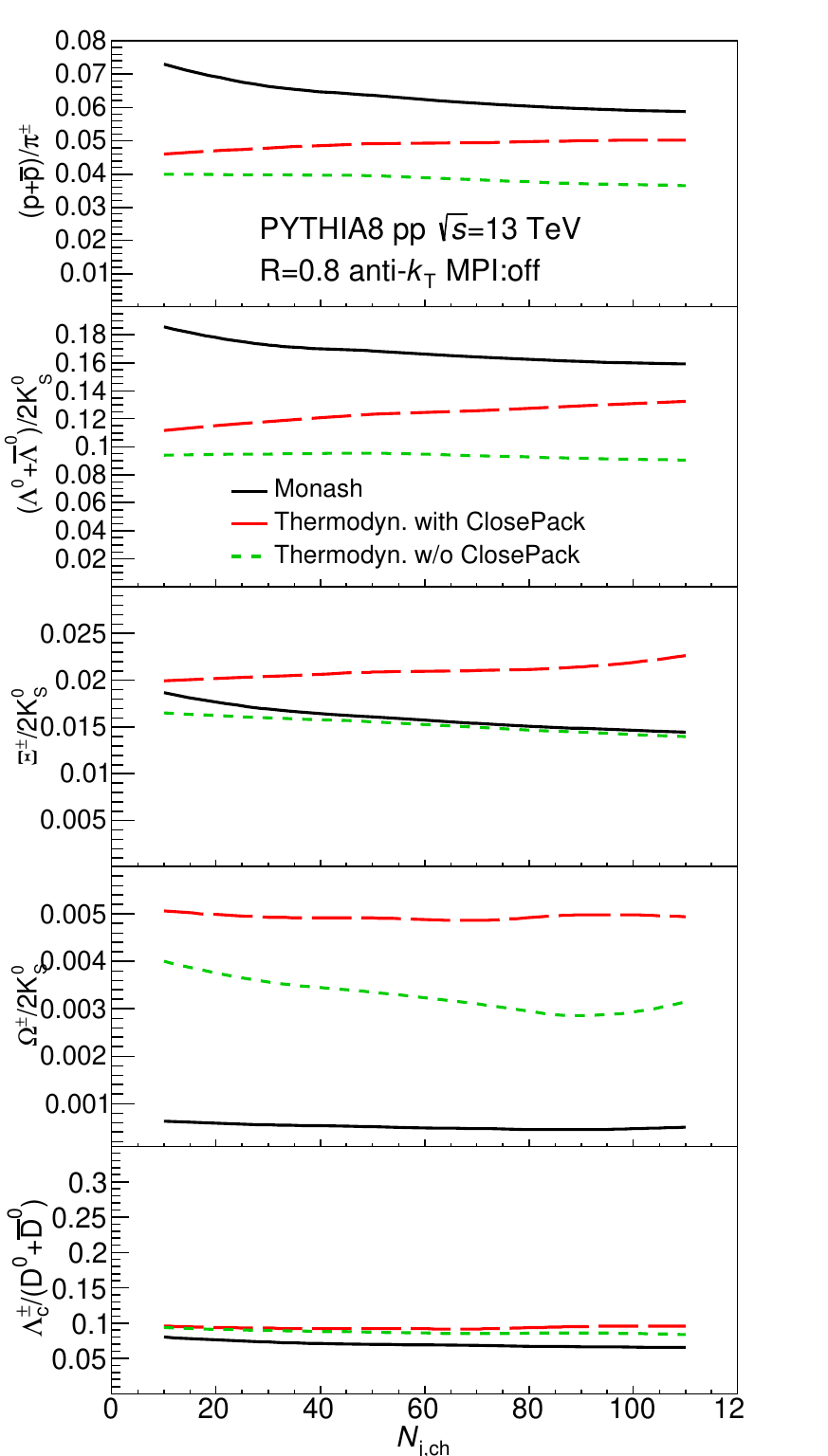}%
	\caption{Baryon-to-meson ratios as a function of the jet charged-constituents multiplicity. Results for pp collisions at $\sqrt{s}=13$\,TeV simulated with different \py settings. The different colour-reconnection schemes (left) and the effect of thermodynamical string fragmentation with and without close packing (right) are compared to the Monash tune. MPI is off in all cases.} 
	\label{fig4}%
\end{figure*}

Figure~\ref{fig4} shows the baryon-to-meson ratios integrated over \jT, as a function of \Nchj. A notable feature is that the standard Lund string fragmentation (Monash, Monash CR-QCD, and CR-BLC mode2 models) predicts higher \pTopi and \LToKzs and lower \XiToKzs and \OmToKzs ratios than the Thermodyn.~CR-QCD model in the full \Nchj range. The standard Lund model predicts either decreasing or nearly constant particle ratios with increasing \Nchj. In contrast, the Thermodyn.~CR-QCD model shows a continuous increase in \pTopi, \LToKzs, and \XiToKzs with increasing \Nchj. This indicates a kind of baryon enhancement for jet constituents. However, for \OmToKzs, the Thermodyn.~CR-QCD model predicts a ratio that is 5 times higher than that predicted by other models, and a decreasing trend with increasing jet multiplicity, which aligns with the findings reported by authors of the thermodynamical string fragmentation model~\cite{Fischer:2016zzs}. In this model, heavier hadrons tend to be produced at large \pt values (in the jet), which might explain the difference from the standard Lund fragmentation model. The same paper notes that the $\Omega$ yield normalised to the charged-pion yield decreases with growing multiplicity, likely due to phase-space constraints. Looking at Table~\ref{tab:pythia}, one notices the parameter {\tt StringFlav:BtoMratio} in the Thermodyn.~CR-QCD model, which controls the relative rate of baryon-to-meson production. We changed this parameter and verified that the multiplicity dependence of the baryon-to-meson ratios were not affected by such a variation. 

Finally, the \LcToDz ratio shows an increasing trend with \Nchj for the models including colour-string junctions (Monash~CR-QCD, CR-BLC~mode2 and Thermodyn.~CR-QCD models), while there is an opposite trend for Monash. It is worth mentioning that ALICE data exhibit a \pT-integrated \LcToDz ratio that is multiplicity independent. Thus, intra-jet studies in the data could be highly valuable for testing different baryon production models and contribute to a deeper understanding of the QGP-like effects observed in small systems.  

It is worth noting that the Thermodyn.~CR-QCD model produces baryon enhancement relative to mesons (except for the $\Omega$ baryon), and also predicts a \LcToDz ratio that starts from $\approx0.11$ (close to the e$^{+}$e$^{-}$ limit) followed by an increase with increasing \Nchj. Then the question is whether these effects have to do with the CR model and/or the close-packing mechanism. The plot on the right-hand side of Fig.~\ref{fig4} shows the baryon-to-meson ratios integrated over \jT, as a function of \Nchj for the Thermodyn. model, where no colour-string junctions are considered. Simulations including the close-packing mechanism exhibit baryon enhancement relative to mesons for \pTopi, \LToKzs, \XiToKzs, and \OmToKzs particle ratios.  On the contrary, the enhancement is not observed when the close-packing mechanism is disabled. Therefore, such an enhancement in the light and strange baryon sector is due to thermodynamical string fragmentation and the close-packing mechanism. In contrast, looking at the heavy flavour sector, the \LcToDz ratio is nearly flat in Thermodyn. model suggesting the importance of colour-string junctions in the production of heavy baryons.

\section{\label{sec:5}Conclusions}
This paper reports on intra-jet identified-hadron production as a function of the jet charged-constituent multiplicity (\Nchj) in \pp collisions at $\sqrt{s}=13$\,TeV simulated with \py~8. The jet frame is used to define the kinematics of jet constituents: \jT is the momentum component perpendicular to the jet axis, \etas and \phis are the pseudorapidity and azimuthal angle coordinates with respect to the jet axis, respectively. The multiparton interaction mechanism is turned off. Given that only charged particles are considered to calculate \Nchj, the light-flavour baryon yields are normalised to the ${\rm K}^{0}_{\rm S}$ yield. Thus, no bias in the neutral-to-charged particle ratio affects the results.

The pp collisions are simulated using the Monash tune, which implements the MPI-based CR model. The effect of colour-string junctions is studied by means of a comparison between the Monash tune and the so-called Monash CR-QCD model. Other studies were performed that included variations of the Monash tune, including thermodynamical string fragmentation with MPI-based and QCD-based CR models named Thermodyn. and Thermodyn.~CR-QCD models, respectively. Finally, the model CR-BLC mode2, which is a tuned version of PYTHIA~8 including colour-string junctions, was also included in the study.

\begin{itemize}
    \item The \Nchj and \etas distributions are only slightly or not affected by the choice of the \py settings.
    
    \item For all \py models the proton-to-pion and the $\Lambda^0$-to-\Kzs ratios show low-high multiplicity grouping trend for $j_{\rm T}>5$\,GeV/$c$, possibly reflecting the flavour of the jet-initiating parton. The ratios show a bump structure that shifts towards higher \jT values with increasing \Nchj. This effect is reminiscent of the radial flow-like effects observed in the \pt spectra as a function of the inclusive charged particle multiplicity. These features, except the multiplicity grouping, are also seen in the $\Xi^\pm$-to-\Kzs particle ratios. The presence of the bump is not obvious in the case of the multistrange $\Omega^\pm$-to-\Kzs ratio, except in the thermodynamical string fragmentation model, where a strong enhancement of  $\Omega^\pm$ is observed compared to other models. 
    
    \item \LcToDz as a function of \jT is little or not multiplicity dependent for Monash tune. However, colour-string junctions (Monash CR-QCD) produce an increase in the \LcToDz ratio at low \jT with increasing \Nchj. The effect gets enhanced for the tune CR-BLC mode2, which also includes colour-string junctions. The size of the effects for the later model is consistent with that produced by Termodyn. CR-QCD model. This effect is similar to the multiplicity-dependent \LcToDz as a function \pt reported by ALICE. For the lowest \Nchj class, the ratio is nearly flat and is consistent with the corresponding ratios measured in ${\rm e}^{+}{\rm e}^{-}$ collisions at LEP. 
    
    \item Regarding the \jT-integrated baryon-to-meson particle ratios for the light-flavoured sector, the Thermodyn. model with string density effects (close-packing mechanism) produces  ratios that increase with \Nchj. On the other hand, the \LcToDz ratio shows an increasing trend with \Nchj for models including colour-string junctions. The ratio departs from $\approx0.11$ (consistent with e$^{+}$e$^{-}$ limit) and increases with \Nchj. These findings encourage intra-jet studies to be relevant in covering the multiplicity gap between e$^{+}$e$^{-}$ and pp collisions.    
\end{itemize}
The results presented in this paper provide new observables that can be used to test thermodynamics-inspired fragmentation models, as well as the role of colour-string junctions in the heavy-flavoured hadron production from small to large systems. 

\begin{table*}[ht]
\centering
\begin{tabular}{cccccc}
\toprule
\multirow{2}{*}{PYTHIA 8 Parameter} & \multirow{2}{*}{Monash} & Monash & CR-BLC & \multirow{2}{*}{Thermodyn.} & Thermodyn. \\ 
    & &  CR-QCD  & (mode 2) &      & CR-QCD \\ 

\midrule
StringPT:sigma & 0.335 & 0.335 & 0.335 & 0.335 & 0.335 \\
StringPT:closePacking & off & off & off & on & on \\
StringPT:thermalModel & off & off & off & on & on \\
StringPT:temperature & & & & 0.21 & 0.21 \\ 
StringPT:expNSP & & & & 0.13 & 0.13 \\
\hline
StringZ:aLund & 0.68 & 0.68 & 0.36 & 0.68 & 0.68 \\
StringZ:bLund & 0.98 & 0.98 & 0.56 & 0.98 & 0.68 \\
\hline
StringFlav:BtoMratio & & & & 0.357 & 0.357 \\
StringFlav:StrangeSuppression &  & &  & 0.357 & 0.357 \\
StringFlav:probQQtoQ & 0.081 & 0.081 & 0.078 & &  \\
StringFlav:ProbStoUD & 0.217 & 0.217 & 0.2 & & \\
\multirow{4}{*}{StringFlav:probQQ1toQQ0join\scalebox{1.3}{\Bigg\{ }} & 0.5 & 0.5 & 0.0275 & 0.5 & 0.5 \\ 
                                                     & 0.7 & 0.7 & 0.0275 & 0.7 & 0.7 \\ 
                                                     & 0.9 & 0.9 & 0.0275 & 0.9 & 0.9 \\ 
                                                     & 1.0 & 1.0 & 0.0275 & 1.0 & 1.0 \\ 
\hline
BeamRemnants:remnantMode & 0 & 1 & 1 & 0 & 1 \\
BeamRemnants:saturation & & 5 & 5 & & 5 \\
\hline
ColourReconnection:mode & 0 & 1 & 1 & 0 & 1 \\
ColourReconnection:allowDoubleJunRem & on & on & off & on & on \\
ColourReconnection:m0 & 0.3 & 0.3 & 0.3 & 0.3 & 0.3 \\
ColourReconnection:allowJunctions & & on & on & &  on \\
ColourReconnection:junctionCorrection & & 1.20 & 1.20 & & 1.20 \\
ColourReconnection:timeDilationMode & & 2 & 2 & & 2 \\
ColourReconnection:timeDilationPar & & 0.18 & 0.18 & & 0.18 \\

\bottomrule
\end{tabular}
\caption{List of the relevant parameters of the different PYTHIA 8 settings used in the current work. Also see Refs.~\cite{Skands:2014pea,Christiansen:2015yqa,Fischer:2016zzs}}
\label{tab:pythia}
\end{table*}

\section*{Acknowledgements}
Authors acknowledge the useful discussions with Torbj{\"o}rn Sj{\"o}strand. R.~V. acknowledges the support by the Hungarian National Research, Development and Innovation Office (NKFIH) under the contract numbers OTKA FK131979 and 2021-4.1.2-NEMZ\_KI-2022-00034. A.~O. acknowledges the support by DGAPA-UNAM under the grants PAPIIT No. IG100524 and PAPIME No. PE100124.

\nocite{*}

\bibliography{injetdyn}

\begin{thebibliography}{38}%
\makeatletter
\providecommand \@ifxundefined [1]{%
 \@ifx{#1\undefined}
}%
\providecommand \@ifnum [1]{%
 \ifnum #1\expandafter \@firstoftwo
 \else \expandafter \@secondoftwo
 \fi
}%
\providecommand \@ifx [1]{%
 \ifx #1\expandafter \@firstoftwo
 \else \expandafter \@secondoftwo
 \fi
}%
\providecommand \natexlab [1]{#1}%
\providecommand \enquote  [1]{``#1''}%
\providecommand \bibnamefont  [1]{#1}%
\providecommand \bibfnamefont [1]{#1}%
\providecommand \citenamefont [1]{#1}%
\providecommand \href@noop [0]{\@secondoftwo}%
\providecommand \href [0]{\begingroup \@sanitize@url \@href}%
\providecommand \@href[1]{\@@startlink{#1}\@@href}%
\providecommand \@@href[1]{\endgroup#1\@@endlink}%
\providecommand \@sanitize@url [0]{\catcode `\\12\catcode `\$12\catcode
  `\&12\catcode `\#12\catcode `\^12\catcode `\_12\catcode `\%12\relax}%
\providecommand \@@startlink[1]{}%
\providecommand \@@endlink[0]{}%
\providecommand \url  [0]{\begingroup\@sanitize@url \@url }%
\providecommand \@url [1]{\endgroup\@href {#1}{\urlprefix }}%
\providecommand \urlprefix  [0]{URL }%
\providecommand \Eprint [0]{\href }%
\providecommand \doibase [0]{https://doi.org/}%
\providecommand \selectlanguage [0]{\@gobble}%
\providecommand \bibinfo  [0]{\@secondoftwo}%
\providecommand \bibfield  [0]{\@secondoftwo}%
\providecommand \translation [1]{[#1]}%
\providecommand \BibitemOpen [0]{}%
\providecommand \bibitemStop [0]{}%
\providecommand \bibitemNoStop [0]{.\EOS\space}%
\providecommand \EOS [0]{\spacefactor3000\relax}%
\providecommand \BibitemShut  [1]{\csname bibitem#1\endcsname}%
\let\auto@bib@innerbib\@empty
\bibitem [{\citenamefont {Busza}\ \emph {et~al.}(2018)\citenamefont {Busza},
  \citenamefont {Rajagopal},\ and\ \citenamefont {van~der
  Schee}}]{Busza:2018rrf}%
  \BibitemOpen
  \bibfield  {author} {\bibinfo {author} {\bibfnamefont {W.}~\bibnamefont
  {Busza}}, \bibinfo {author} {\bibfnamefont {K.}~\bibnamefont {Rajagopal}},\
  and\ \bibinfo {author} {\bibfnamefont {W.}~\bibnamefont {van~der Schee}},\
  }\bibfield  {title} {\bibinfo {title} {{Heavy Ion Collisions: The Big
  Picture, and the Big Questions}},\ }\href
  {https://doi.org/10.1146/annurev-nucl-101917-020852} {\bibfield  {journal}
  {\bibinfo  {journal} {Ann. Rev. Nucl. Part. Sci.}\ }\textbf {\bibinfo
  {volume} {68}},\ \bibinfo {pages} {339} (\bibinfo {year} {2018})},\ \Eprint
  {https://arxiv.org/abs/1802.04801} {arXiv:1802.04801 [hep-ph]} \BibitemShut
  {NoStop}%
\bibitem [{\citenamefont {Bala}\ \emph {et~al.}(2016)\citenamefont {Bala},
  \citenamefont {Bautista}, \citenamefont {Bielcikova},\ and\ \citenamefont
  {Ortiz}}]{Bala:2016hlf}%
  \BibitemOpen
  \bibfield  {author} {\bibinfo {author} {\bibfnamefont {R.}~\bibnamefont
  {Bala}}, \bibinfo {author} {\bibfnamefont {I.}~\bibnamefont {Bautista}},
  \bibinfo {author} {\bibfnamefont {J.}~\bibnamefont {Bielcikova}},\ and\
  \bibinfo {author} {\bibfnamefont {A.}~\bibnamefont {Ortiz}},\ }\bibfield
  {title} {\bibinfo {title} {{Heavy-ion physics at the LHC: Review of Run I
  results}},\ }\href {https://doi.org/10.1142/S0218301316420064} {\bibfield
  {journal} {\bibinfo  {journal} {Int. J. Mod. Phys. E}\ }\textbf {\bibinfo
  {volume} {25}},\ \bibinfo {pages} {1642006} (\bibinfo {year} {2016})},\
  \Eprint {https://arxiv.org/abs/1605.03939} {arXiv:1605.03939 [hep-ex]}
  \BibitemShut {NoStop}%
\bibitem [{\citenamefont {Acharya}\ \emph
  {et~al.}(2022{\natexlab{a}})\citenamefont {Acharya} \emph
  {et~al.}}]{ALICE:2022wpn}%
  \BibitemOpen
  \bibfield  {author} {\bibinfo {author} {\bibfnamefont {S.}~\bibnamefont
  {Acharya}} \emph {et~al.} (\bibinfo {collaboration} {ALICE}),\ }\bibfield
  {title} {\bibinfo {title} {{The ALICE experiment -- A journey through QCD}},\
  }\href@noop {} {\  (\bibinfo {year} {2022}{\natexlab{a}})},\ \Eprint
  {https://arxiv.org/abs/2211.04384} {arXiv:2211.04384 [nucl-ex]} \BibitemShut
  {NoStop}%
\bibitem [{\citenamefont {Acharya}\ \emph {et~al.}(2020)\citenamefont {Acharya}
  \emph {et~al.}}]{ALICE:2020fuk}%
  \BibitemOpen
  \bibfield  {author} {\bibinfo {author} {\bibfnamefont {S.}~\bibnamefont
  {Acharya}} \emph {et~al.} (\bibinfo {collaboration} {ALICE}),\ }\bibfield
  {title} {\bibinfo {title} {{Future high-energy pp programme with ALICE}},\
  }\href@noop {} {\bibfield  {journal} {\bibinfo  {journal}
  {ALICE-PUBLIC-2020-005}\ } (\bibinfo {year} {2020})}\BibitemShut {NoStop}%
\bibitem [{\citenamefont {Acharya}\ \emph
  {et~al.}(2023{\natexlab{a}})\citenamefont {Acharya} \emph
  {et~al.}}]{ALICE:2022qxg}%
  \BibitemOpen
  \bibfield  {author} {\bibinfo {author} {\bibfnamefont {S.}~\bibnamefont
  {Acharya}} \emph {et~al.} (\bibinfo {collaboration} {ALICE}),\ }\bibfield
  {title} {\bibinfo {title} {{Study of charged particle production at high
  $p_{\rm T}$ using event topology in pp, p--Pb and Pb--Pb collisions at
  $\sqrt{s_{\rm NN}}=5.02$ TeV}},\ }\href
  {https://doi.org/10.1016/j.physletb.2022.137649} {\bibfield  {journal}
  {\bibinfo  {journal} {Phys. Lett. B}\ }\textbf {\bibinfo {volume} {843}},\
  \bibinfo {pages} {137649} (\bibinfo {year} {2023}{\natexlab{a}})},\ \Eprint
  {https://arxiv.org/abs/2204.10157} {arXiv:2204.10157 [nucl-ex]} \BibitemShut
  {NoStop}%
\bibitem [{\citenamefont {Acharya}\ \emph
  {et~al.}(2024{\natexlab{a}})\citenamefont {Acharya} \emph
  {et~al.}}]{ALICE:2023plt}%
  \BibitemOpen
  \bibfield  {author} {\bibinfo {author} {\bibfnamefont {S.}~\bibnamefont
  {Acharya}} \emph {et~al.} (\bibinfo {collaboration} {ALICE}),\ }\bibfield
  {title} {\bibinfo {title} {{Search for jet quenching effects in
  high-multiplicity pp collisions at $ \sqrt{s} $ = 13 TeV via di-jet
  acoplanarity}},\ }\href {https://doi.org/10.1007/JHEP05(2024)229} {\bibfield
  {journal} {\bibinfo  {journal} {JHEP}\ }\textbf {\bibinfo {volume} {05}},\
  \bibinfo {pages} {229}},\ \Eprint {https://arxiv.org/abs/2309.03788}
  {arXiv:2309.03788 [hep-ex]} \BibitemShut {NoStop}%
\bibitem [{\citenamefont {Acharya}\ \emph
  {et~al.}(2024{\natexlab{b}})\citenamefont {Acharya} \emph
  {et~al.}}]{ALICE:2023ulm}%
  \BibitemOpen
  \bibfield  {author} {\bibinfo {author} {\bibfnamefont {S.}~\bibnamefont
  {Acharya}} \emph {et~al.} (\bibinfo {collaboration} {ALICE}),\ }\bibfield
  {title} {\bibinfo {title} {{Emergence of long-range angular correlations in
  low-multiplicity proton-proton Collisions}},\ }\href
  {https://doi.org/10.1103/PhysRevLett.132.172302} {\bibfield  {journal}
  {\bibinfo  {journal} {Phys. Rev. Lett.}\ }\textbf {\bibinfo {volume} {132}},\
  \bibinfo {pages} {172302} (\bibinfo {year} {2024}{\natexlab{b}})},\ \Eprint
  {https://arxiv.org/abs/2311.14357} {arXiv:2311.14357 [nucl-ex]} \BibitemShut
  {NoStop}%
\bibitem [{\citenamefont {Aad}\ \emph {et~al.}(2021)\citenamefont {Aad} \emph
  {et~al.}}]{ATLAS:2021jhn}%
  \BibitemOpen
  \bibfield  {author} {\bibinfo {author} {\bibfnamefont {G.}~\bibnamefont
  {Aad}} \emph {et~al.} (\bibinfo {collaboration} {ATLAS}),\ }\bibfield
  {title} {\bibinfo {title} {{Two-particle azimuthal correlations in
  photonuclear ultraperipheral Pb--Pb collisions at $\sqrt{s_{\rm NN}}=5.02$
  TeV with ATLAS}},\ }\href {https://doi.org/10.1103/PhysRevC.104.014903}
  {\bibfield  {journal} {\bibinfo  {journal} {Phys. Rev. C}\ }\textbf {\bibinfo
  {volume} {104}},\ \bibinfo {pages} {014903} (\bibinfo {year} {2021})},\
  \Eprint {https://arxiv.org/abs/2101.10771} {arXiv:2101.10771 [nucl-ex]}
  \BibitemShut {NoStop}%
\bibitem [{\citenamefont {Hayrapetyan}\ \emph {et~al.}(2024)\citenamefont
  {Hayrapetyan} \emph {et~al.}}]{CMS:2023iam}%
  \BibitemOpen
  \bibfield  {author} {\bibinfo {author} {\bibfnamefont {A.}~\bibnamefont
  {Hayrapetyan}} \emph {et~al.} (\bibinfo {collaboration} {CMS}),\ }\bibfield
  {title} {\bibinfo {title} {{Observation of Enhanced Long-Range Elliptic
  Anisotropies Inside High-Multiplicity Jets in pp Collisions at
  s=13\,\,TeV}},\ }\href {https://doi.org/10.1103/PhysRevLett.133.142301}
  {\bibfield  {journal} {\bibinfo  {journal} {Phys. Rev. Lett.}\ }\textbf
  {\bibinfo {volume} {133}},\ \bibinfo {pages} {142301} (\bibinfo {year}
  {2024})},\ \Eprint {https://arxiv.org/abs/2312.17103} {arXiv:2312.17103
  [hep-ex]} \BibitemShut {NoStop}%
\bibitem [{\citenamefont {Werthmann}\ \emph {et~al.}(2024)\citenamefont
  {Werthmann}, \citenamefont {Ambrus},\ and\ \citenamefont
  {Schlichting}}]{Werthmann:2023dvl}%
  \BibitemOpen
  \bibfield  {author} {\bibinfo {author} {\bibfnamefont {C.}~\bibnamefont
  {Werthmann}}, \bibinfo {author} {\bibfnamefont {V.~E.}\ \bibnamefont
  {Ambrus}},\ and\ \bibinfo {author} {\bibfnamefont {S.}~\bibnamefont
  {Schlichting}},\ }\bibfield  {title} {\bibinfo {title} {{System size
  dependence of pre-equilibrium and applicability of hydrodynamics in heavy-ion
  collisions}},\ }\href {https://doi.org/10.22323/1.438.0048} {\bibfield
  {journal} {\bibinfo  {journal} {PoS}\ }\textbf {\bibinfo {volume}
  {HardProbes2023}},\ \bibinfo {pages} {048} (\bibinfo {year} {2024})},\
  \Eprint {https://arxiv.org/abs/2307.08306} {arXiv:2307.08306 [hep-ph]}
  \BibitemShut {NoStop}%
\bibitem [{\citenamefont {d'Enterria}\ \emph {et~al.}(2010)\citenamefont
  {d'Enterria}, \citenamefont {Eyyubova}, \citenamefont {Korotkikh},
  \citenamefont {Lokhtin}, \citenamefont {Petrushanko}, \citenamefont
  {Sarycheva},\ and\ \citenamefont {Snigirev}}]{dEnterria:2010xip}%
  \BibitemOpen
  \bibfield  {author} {\bibinfo {author} {\bibfnamefont {D.}~\bibnamefont
  {d'Enterria}}, \bibinfo {author} {\bibfnamefont {G.~K.}\ \bibnamefont
  {Eyyubova}}, \bibinfo {author} {\bibfnamefont {V.~L.}\ \bibnamefont
  {Korotkikh}}, \bibinfo {author} {\bibfnamefont {I.~P.}\ \bibnamefont
  {Lokhtin}}, \bibinfo {author} {\bibfnamefont {S.~V.}\ \bibnamefont
  {Petrushanko}}, \bibinfo {author} {\bibfnamefont {L.~I.}\ \bibnamefont
  {Sarycheva}},\ and\ \bibinfo {author} {\bibfnamefont {A.~M.}\ \bibnamefont
  {Snigirev}},\ }\bibfield  {title} {\bibinfo {title} {{Estimates of hadron
  azimuthal anisotropy from multiparton interactions in proton-proton
  collisions at $\sqrt{s}=14$ TeV}},\ }\href
  {https://doi.org/10.1140/epjc/s10052-009-1232-7} {\bibfield  {journal}
  {\bibinfo  {journal} {Eur. Phys. J. C}\ }\textbf {\bibinfo {volume} {66}},\
  \bibinfo {pages} {173} (\bibinfo {year} {2010})},\ \Eprint
  {https://arxiv.org/abs/0910.3029} {arXiv:0910.3029 [hep-ph]} \BibitemShut
  {NoStop}%
\bibitem [{\citenamefont {Bierlich}\ \emph {et~al.}(2015)\citenamefont
  {Bierlich}, \citenamefont {Gustafson}, \citenamefont {L\"onnblad},\ and\
  \citenamefont {Tarasov}}]{Bierlich:2014xba}%
  \BibitemOpen
  \bibfield  {author} {\bibinfo {author} {\bibfnamefont {C.}~\bibnamefont
  {Bierlich}}, \bibinfo {author} {\bibfnamefont {G.}~\bibnamefont {Gustafson}},
  \bibinfo {author} {\bibfnamefont {L.}~\bibnamefont {L\"onnblad}},\ and\
  \bibinfo {author} {\bibfnamefont {A.}~\bibnamefont {Tarasov}},\ }\bibfield
  {title} {\bibinfo {title} {{Effects of overlapping strings in pp
  collisions}},\ }\href {https://doi.org/10.1007/JHEP03(2015)148} {\bibfield
  {journal} {\bibinfo  {journal} {JHEP}\ }\textbf {\bibinfo {volume} {03}},\
  \bibinfo {pages} {148}},\ \Eprint {https://arxiv.org/abs/1412.6259}
  {arXiv:1412.6259 [hep-ph]} \BibitemShut {NoStop}%
\bibitem [{\citenamefont {Bierlich}\ \emph {et~al.}(2018)\citenamefont
  {Bierlich}, \citenamefont {Gustafson}, \citenamefont {L\"onnblad},\ and\
  \citenamefont {Shah}}]{Bierlich:2018xfw}%
  \BibitemOpen
  \bibfield  {author} {\bibinfo {author} {\bibfnamefont {C.}~\bibnamefont
  {Bierlich}}, \bibinfo {author} {\bibfnamefont {G.}~\bibnamefont {Gustafson}},
  \bibinfo {author} {\bibfnamefont {L.}~\bibnamefont {L\"onnblad}},\ and\
  \bibinfo {author} {\bibfnamefont {H.}~\bibnamefont {Shah}},\ }\bibfield
  {title} {\bibinfo {title} {{The Angantyr model for heavy-ion collisions in
  PYTHIA 8}},\ }\href {https://doi.org/10.1007/JHEP10(2018)134} {\bibfield
  {journal} {\bibinfo  {journal} {JHEP}\ }\textbf {\bibinfo {volume} {10}},\
  \bibinfo {pages} {134}},\ \Eprint {https://arxiv.org/abs/1806.10820}
  {arXiv:1806.10820 [hep-ph]} \BibitemShut {NoStop}%
\bibitem [{\citenamefont {Ortiz}\ \emph {et~al.}(2024)\citenamefont {Ortiz},
  \citenamefont {Bencedi},\ and\ \citenamefont {Fan}}]{Ortiz:2024ndh}%
  \BibitemOpen
  \bibfield  {author} {\bibinfo {author} {\bibfnamefont {A.}~\bibnamefont
  {Ortiz}}, \bibinfo {author} {\bibfnamefont {G.}~\bibnamefont {Bencedi}},\
  and\ \bibinfo {author} {\bibfnamefont {F.}~\bibnamefont {Fan}},\ }\bibfield
  {title} {\bibinfo {title} {{Flattenicity as centrality estimator in
  p\textendash{}Pb collisions simulated with PYTHIA/Angantyr}},\ }\href
  {https://doi.org/10.1088/1361-6471/ad8249} {\bibfield  {journal} {\bibinfo
  {journal} {J. Phys. G}\ }\textbf {\bibinfo {volume} {51}},\ \bibinfo {pages}
  {125003} (\bibinfo {year} {2024})},\ \Eprint
  {https://arxiv.org/abs/2407.07724} {arXiv:2407.07724 [nucl-ex]} \BibitemShut
  {NoStop}%
\bibitem [{\citenamefont {Baty}\ \emph {et~al.}(2023)\citenamefont {Baty},
  \citenamefont {Gardner},\ and\ \citenamefont {Li}}]{Baty:2021ugw}%
  \BibitemOpen
  \bibfield  {author} {\bibinfo {author} {\bibfnamefont {A.}~\bibnamefont
  {Baty}}, \bibinfo {author} {\bibfnamefont {P.}~\bibnamefont {Gardner}},\ and\
  \bibinfo {author} {\bibfnamefont {W.}~\bibnamefont {Li}},\ }\bibfield
  {title} {\bibinfo {title} {{Novel observables for exploring QCD collective
  evolution and quantum entanglement within individual jets}},\ }\href
  {https://doi.org/10.1103/PhysRevC.107.064908} {\bibfield  {journal} {\bibinfo
   {journal} {Phys. Rev. C}\ }\textbf {\bibinfo {volume} {107}},\ \bibinfo
  {pages} {064908} (\bibinfo {year} {2023})},\ \Eprint
  {https://arxiv.org/abs/2104.11735} {arXiv:2104.11735 [hep-ph]} \BibitemShut
  {NoStop}%
\bibitem [{\citenamefont {Ortiz}\ \emph {et~al.}(2013)\citenamefont {Ortiz},
  \citenamefont {Christiansen}, \citenamefont {Cuautle}, \citenamefont
  {Maldonado},\ and\ \citenamefont {Pai\'c}}]{OrtizVelasquez:2013ofg}%
  \BibitemOpen
  \bibfield  {author} {\bibinfo {author} {\bibfnamefont {A.}~\bibnamefont
  {Ortiz}}, \bibinfo {author} {\bibfnamefont {P.}~\bibnamefont {Christiansen}},
  \bibinfo {author} {\bibfnamefont {E.}~\bibnamefont {Cuautle}}, \bibinfo
  {author} {\bibfnamefont {I.}~\bibnamefont {Maldonado}},\ and\ \bibinfo
  {author} {\bibfnamefont {G.}~\bibnamefont {Pai\'c}},\ }\bibfield  {title}
  {\bibinfo {title} {{Color reconnection and flowlike patterns in pp
  collisions}},\ }\href {https://doi.org/10.1103/PhysRevLett.111.042001}
  {\bibfield  {journal} {\bibinfo  {journal} {Phys. Rev. Lett.}\ }\textbf
  {\bibinfo {volume} {111}},\ \bibinfo {pages} {042001} (\bibinfo {year}
  {2013})},\ \Eprint {https://arxiv.org/abs/1303.6326} {arXiv:1303.6326
  [hep-ph]} \BibitemShut {NoStop}%
\bibitem [{\citenamefont {Fischer}\ and\ \citenamefont
  {Sj\"ostrand}(2017)}]{Fischer:2016zzs}%
  \BibitemOpen
  \bibfield  {author} {\bibinfo {author} {\bibfnamefont {N.}~\bibnamefont
  {Fischer}}\ and\ \bibinfo {author} {\bibfnamefont {T.}~\bibnamefont
  {Sj\"ostrand}},\ }\bibfield  {title} {\bibinfo {title} {{Thermodynamical
  string fragmentation}},\ }\href {https://doi.org/10.1007/JHEP01(2017)140}
  {\bibfield  {journal} {\bibinfo  {journal} {JHEP}\ }\textbf {\bibinfo
  {volume} {01}},\ \bibinfo {pages} {140}},\ \Eprint
  {https://arxiv.org/abs/1610.09818} {arXiv:1610.09818 [hep-ph]} \BibitemShut
  {NoStop}%
\bibitem [{\citenamefont {Becattini}(1995)}]{Becattini:1995hr}%
  \BibitemOpen
  \bibfield  {author} {\bibinfo {author} {\bibfnamefont {F.}~\bibnamefont
  {Becattini}},\ }\bibfield  {title} {\bibinfo {title} {{A Thermodynamical
  model of hadron production in e+ e- collisions}},\ }in\ \href@noop {} {\emph
  {\bibinfo {booktitle} {{25th International Symposium on Multiparticle
  Dynamics}}}}\ (\bibinfo {year} {1995})\ pp.\ \bibinfo {pages} {480--490},\
  \Eprint {https://arxiv.org/abs/hep-ph/9511235} {arXiv:hep-ph/9511235}
  \BibitemShut {NoStop}%
\bibitem [{\citenamefont {Sj\"ostrand}\ \emph {et~al.}(2015)\citenamefont
  {Sj\"ostrand}, \citenamefont {Ask}, \citenamefont {Christiansen},
  \citenamefont {Corke}, \citenamefont {Desai}, \citenamefont {Ilten},
  \citenamefont {Mrenna}, \citenamefont {Prestel}, \citenamefont {Rasmussen},\
  and\ \citenamefont {Skands}}]{Sjostrand:2014zea}%
  \BibitemOpen
  \bibfield  {author} {\bibinfo {author} {\bibfnamefont {T.}~\bibnamefont
  {Sj\"ostrand}}, \bibinfo {author} {\bibfnamefont {S.}~\bibnamefont {Ask}},
  \bibinfo {author} {\bibfnamefont {J.~R.}\ \bibnamefont {Christiansen}},
  \bibinfo {author} {\bibfnamefont {R.}~\bibnamefont {Corke}}, \bibinfo
  {author} {\bibfnamefont {N.}~\bibnamefont {Desai}}, \bibinfo {author}
  {\bibfnamefont {P.}~\bibnamefont {Ilten}}, \bibinfo {author} {\bibfnamefont
  {S.}~\bibnamefont {Mrenna}}, \bibinfo {author} {\bibfnamefont
  {S.}~\bibnamefont {Prestel}}, \bibinfo {author} {\bibfnamefont {C.~O.}\
  \bibnamefont {Rasmussen}},\ and\ \bibinfo {author} {\bibfnamefont {P.~Z.}\
  \bibnamefont {Skands}},\ }\bibfield  {title} {\bibinfo {title} {{An
  introduction to PYTHIA 8.2}},\ }\href
  {https://doi.org/10.1016/j.cpc.2015.01.024} {\bibfield  {journal} {\bibinfo
  {journal} {Comput. Phys. Commun.}\ }\textbf {\bibinfo {volume} {191}},\
  \bibinfo {pages} {159} (\bibinfo {year} {2015})},\ \Eprint
  {https://arxiv.org/abs/1410.3012} {arXiv:1410.3012 [hep-ph]} \BibitemShut
  {NoStop}%
\bibitem [{\citenamefont {Andersson}\ \emph {et~al.}(1983)\citenamefont
  {Andersson}, \citenamefont {Gustafson}, \citenamefont {Ingelman},\ and\
  \citenamefont {Sj\"ostrand}}]{Andersson:1983ia}%
  \BibitemOpen
  \bibfield  {author} {\bibinfo {author} {\bibfnamefont {B.}~\bibnamefont
  {Andersson}}, \bibinfo {author} {\bibfnamefont {G.}~\bibnamefont
  {Gustafson}}, \bibinfo {author} {\bibfnamefont {G.}~\bibnamefont
  {Ingelman}},\ and\ \bibinfo {author} {\bibfnamefont {T.}~\bibnamefont
  {Sj\"ostrand}},\ }\bibfield  {title} {\bibinfo {title} {{Parton fragmentation
  and string dynamics}},\ }\href {https://doi.org/10.1016/0370-1573(83)90080-7}
  {\bibfield  {journal} {\bibinfo  {journal} {Phys. Rept.}\ }\textbf {\bibinfo
  {volume} {97}},\ \bibinfo {pages} {31} (\bibinfo {year} {1983})}\BibitemShut
  {NoStop}%
\bibitem [{\citenamefont {Skands}\ \emph {et~al.}(2014)\citenamefont {Skands},
  \citenamefont {Carrazza},\ and\ \citenamefont {Rojo}}]{Skands:2014pea}%
  \BibitemOpen
  \bibfield  {author} {\bibinfo {author} {\bibfnamefont {P.}~\bibnamefont
  {Skands}}, \bibinfo {author} {\bibfnamefont {S.}~\bibnamefont {Carrazza}},\
  and\ \bibinfo {author} {\bibfnamefont {J.}~\bibnamefont {Rojo}},\ }\bibfield
  {title} {\bibinfo {title} {{Tuning PYTHIA 8.1: the Monash 2013 Tune}},\
  }\href {https://doi.org/10.1140/epjc/s10052-014-3024-y} {\bibfield  {journal}
  {\bibinfo  {journal} {Eur. Phys. J. C}\ }\textbf {\bibinfo {volume} {74}},\
  \bibinfo {pages} {3024} (\bibinfo {year} {2014})},\ \Eprint
  {https://arxiv.org/abs/1404.5630} {arXiv:1404.5630 [hep-ph]} \BibitemShut
  {NoStop}%
\bibitem [{\citenamefont {Christiansen}\ and\ \citenamefont
  {Skands}(2015)}]{Christiansen:2015yqa}%
  \BibitemOpen
  \bibfield  {author} {\bibinfo {author} {\bibfnamefont {J.~R.}\ \bibnamefont
  {Christiansen}}\ and\ \bibinfo {author} {\bibfnamefont {P.~Z.}\ \bibnamefont
  {Skands}},\ }\bibfield  {title} {\bibinfo {title} {{String formation beyond
  leading colour}},\ }\href {https://doi.org/10.1007/JHEP08(2015)003}
  {\bibfield  {journal} {\bibinfo  {journal} {JHEP}\ }\textbf {\bibinfo
  {volume} {08}},\ \bibinfo {pages} {003}},\ \Eprint
  {https://arxiv.org/abs/1505.01681} {arXiv:1505.01681 [hep-ph]} \BibitemShut
  {NoStop}%
\bibitem [{\citenamefont {Bierlich}\ \emph {et~al.}(2016)\citenamefont
  {Bierlich}, \citenamefont {Gustafson},\ and\ \citenamefont
  {L\"onnblad}}]{Bierlich:2016vgw}%
  \BibitemOpen
  \bibfield  {author} {\bibinfo {author} {\bibfnamefont {C.}~\bibnamefont
  {Bierlich}}, \bibinfo {author} {\bibfnamefont {G.}~\bibnamefont
  {Gustafson}},\ and\ \bibinfo {author} {\bibfnamefont {L.}~\bibnamefont
  {L\"onnblad}},\ }\bibfield  {title} {\bibinfo {title} {{A shoving model for
  collectivity in hadronic collisions}},\ }\href@noop {} {\  (\bibinfo {year}
  {2016})},\ \Eprint {https://arxiv.org/abs/1612.05132} {arXiv:1612.05132
  [hep-ph]} \BibitemShut {NoStop}%
\bibitem [{\citenamefont {Acharya}\ \emph
  {et~al.}(2022{\natexlab{b}})\citenamefont {Acharya} \emph
  {et~al.}}]{ALICE:2021npz}%
  \BibitemOpen
  \bibfield  {author} {\bibinfo {author} {\bibfnamefont {S.}~\bibnamefont
  {Acharya}} \emph {et~al.} (\bibinfo {collaboration} {ALICE}),\ }\bibfield
  {title} {\bibinfo {title} {{Observation of a multiplicity dependence in the
  $p_{\rm T}$-differential charm baryon-to-meson ratios in proton-proton
  collisions at $\sqrt{s}=13$ TeV}},\ }\href
  {https://doi.org/10.1016/j.physletb.2022.137065} {\bibfield  {journal}
  {\bibinfo  {journal} {Phys. Lett. B}\ }\textbf {\bibinfo {volume} {829}},\
  \bibinfo {pages} {137065} (\bibinfo {year} {2022}{\natexlab{b}})},\ \Eprint
  {https://arxiv.org/abs/2111.11948} {arXiv:2111.11948 [nucl-ex]} \BibitemShut
  {NoStop}%
\bibitem [{\citenamefont {Acharya}\ \emph
  {et~al.}(2023{\natexlab{b}})\citenamefont {Acharya} \emph
  {et~al.}}]{ALICE:2023sgl}%
  \BibitemOpen
  \bibfield  {author} {\bibinfo {author} {\bibfnamefont {S.}~\bibnamefont
  {Acharya}} \emph {et~al.} (\bibinfo {collaboration} {ALICE}),\ }\bibfield
  {title} {\bibinfo {title} {{Charm production and fragmentation fractions at
  midrapidity in pp collisions at $ \sqrt{\textrm{s}} $ = 13 TeV}},\ }\href
  {https://doi.org/10.1007/JHEP12(2023)086} {\bibfield  {journal} {\bibinfo
  {journal} {JHEP}\ }\textbf {\bibinfo {volume} {12}},\ \bibinfo {pages}
  {086}},\ \Eprint {https://arxiv.org/abs/2308.04877} {arXiv:2308.04877
  [hep-ex]} \BibitemShut {NoStop}%
\bibitem [{\citenamefont {Sirunyan}\ \emph {et~al.}(2020)\citenamefont
  {Sirunyan} \emph {et~al.}}]{CMS:2019uws}%
  \BibitemOpen
  \bibfield  {author} {\bibinfo {author} {\bibfnamefont {A.~M.}\ \bibnamefont
  {Sirunyan}} \emph {et~al.} (\bibinfo {collaboration} {CMS}),\ }\bibfield
  {title} {\bibinfo {title} {{Production of $\Lambda_\mathrm{c}^+$ baryons in
  proton-proton and lead-lead collisions at $\sqrt{s_\mathrm{NN}}=$ 5.02
  TeV}},\ }\href {https://doi.org/10.1016/j.physletb.2020.135328} {\bibfield
  {journal} {\bibinfo  {journal} {Phys. Lett. B}\ }\textbf {\bibinfo {volume}
  {803}},\ \bibinfo {pages} {135328} (\bibinfo {year} {2020})},\ \Eprint
  {https://arxiv.org/abs/1906.03322} {arXiv:1906.03322 [hep-ex]} \BibitemShut
  {NoStop}%
\bibitem [{\citenamefont {Acharya}\ \emph
  {et~al.}(2021{\natexlab{a}})\citenamefont {Acharya} \emph
  {et~al.}}]{ALICE:2020jsh}%
  \BibitemOpen
  \bibfield  {author} {\bibinfo {author} {\bibfnamefont {S.}~\bibnamefont
  {Acharya}} \emph {et~al.} (\bibinfo {collaboration} {ALICE}),\ }\bibfield
  {title} {\bibinfo {title} {{Production of light-flavor hadrons in pp
  collisions at $\sqrt{s}~=~7\text { and }\sqrt{s} = 13 \, \text { TeV} $}},\
  }\href {https://doi.org/10.1140/epjc/s10052-020-08690-5} {\bibfield
  {journal} {\bibinfo  {journal} {Eur. Phys. J. C}\ }\textbf {\bibinfo {volume}
  {81}},\ \bibinfo {pages} {256} (\bibinfo {year} {2021}{\natexlab{a}})},\
  \Eprint {https://arxiv.org/abs/2005.11120} {arXiv:2005.11120 [nucl-ex]}
  \BibitemShut {NoStop}%
\bibitem [{\citenamefont {Acharya}\ \emph
  {et~al.}(2022{\natexlab{c}})\citenamefont {Acharya} \emph
  {et~al.}}]{ALICE:2021rzj}%
  \BibitemOpen
  \bibfield  {author} {\bibinfo {author} {\bibfnamefont {S.}~\bibnamefont
  {Acharya}} \emph {et~al.} (\bibinfo {collaboration} {ALICE}),\ }\bibfield
  {title} {\bibinfo {title} {{Measurement of Prompt D$^{0}$, $\Lambda_{c}^{+}$,
  and $\Sigma_{c}^{0,++}$(2455) Production in Proton\textendash{}Proton
  Collisions at $\sqrt s$ = 13\,\,TeV}},\ }\href
  {https://doi.org/10.1103/PhysRevLett.128.012001} {\bibfield  {journal}
  {\bibinfo  {journal} {Phys. Rev. Lett.}\ }\textbf {\bibinfo {volume} {128}},\
  \bibinfo {pages} {012001} (\bibinfo {year} {2022}{\natexlab{c}})},\ \Eprint
  {https://arxiv.org/abs/2106.08278} {arXiv:2106.08278 [hep-ex]} \BibitemShut
  {NoStop}%
\bibitem [{\citenamefont {Cacciari}\ \emph {et~al.}(2008)\citenamefont
  {Cacciari}, \citenamefont {Salam},\ and\ \citenamefont
  {Soyez}}]{Cacciari:2008gp}%
  \BibitemOpen
  \bibfield  {author} {\bibinfo {author} {\bibfnamefont {M.}~\bibnamefont
  {Cacciari}}, \bibinfo {author} {\bibfnamefont {G.~P.}\ \bibnamefont
  {Salam}},\ and\ \bibinfo {author} {\bibfnamefont {G.}~\bibnamefont {Soyez}},\
  }\bibfield  {title} {\bibinfo {title} {{The anti-$k_{\rm T}$ jet clustering
  algorithm}},\ }\href {https://doi.org/10.1088/1126-6708/2008/04/063}
  {\bibfield  {journal} {\bibinfo  {journal} {JHEP}\ }\textbf {\bibinfo
  {volume} {04}},\ \bibinfo {pages} {063}},\ \Eprint
  {https://arxiv.org/abs/0802.1189} {arXiv:0802.1189 [hep-ph]} \BibitemShut
  {NoStop}%
\bibitem [{\citenamefont {Acharya}\ \emph {et~al.}(2019)\citenamefont {Acharya}
  \emph {et~al.}}]{ALICE:2018pal}%
  \BibitemOpen
  \bibfield  {author} {\bibinfo {author} {\bibfnamefont {S.}~\bibnamefont
  {Acharya}} \emph {et~al.} (\bibinfo {collaboration} {ALICE}),\ }\bibfield
  {title} {\bibinfo {title} {{Multiplicity dependence of light-flavor hadron
  production in pp collisions at $\sqrt{s}$ = 7 TeV}},\ }\href
  {https://doi.org/10.1103/PhysRevC.99.024906} {\bibfield  {journal} {\bibinfo
  {journal} {Phys. Rev. C}\ }\textbf {\bibinfo {volume} {99}},\ \bibinfo
  {pages} {024906} (\bibinfo {year} {2019})},\ \Eprint
  {https://arxiv.org/abs/1807.11321} {arXiv:1807.11321 [nucl-ex]} \BibitemShut
  {NoStop}%
\bibitem [{\citenamefont {Acharya}\ \emph {et~al.}(2018)\citenamefont {Acharya}
  \emph {et~al.}}]{ALICE:2017thy}%
  \BibitemOpen
  \bibfield  {author} {\bibinfo {author} {\bibfnamefont {S.}~\bibnamefont
  {Acharya}} \emph {et~al.} (\bibinfo {collaboration} {ALICE}),\ }\bibfield
  {title} {\bibinfo {title} {{$\Lambda_{\rm c}^+$ production in pp collisions
  at $\sqrt{s} = 7$ TeV and in p--Pb collisions at $\sqrt{s_{\rm NN}} = 5.02$
  TeV}},\ }\href {https://doi.org/10.1007/JHEP04(2018)108} {\bibfield
  {journal} {\bibinfo  {journal} {JHEP}\ }\textbf {\bibinfo {volume} {04}},\
  \bibinfo {pages} {108}},\ \Eprint {https://arxiv.org/abs/1712.09581}
  {arXiv:1712.09581 [nucl-ex]} \BibitemShut {NoStop}%
\bibitem [{\citenamefont {Acharya}\ \emph
  {et~al.}(2021{\natexlab{b}})\citenamefont {Acharya} \emph
  {et~al.}}]{ALICE:2020wfu}%
  \BibitemOpen
  \bibfield  {author} {\bibinfo {author} {\bibfnamefont {S.}~\bibnamefont
  {Acharya}} \emph {et~al.} (\bibinfo {collaboration} {ALICE}),\ }\bibfield
  {title} {\bibinfo {title} {{$\Lambda^+_c$ production and baryon-to-meson
  ratios in pp and p--Pb collisions at $\sqrt {s_{\rm NN}}$=5.02 TeV at the
  LHC}},\ }\href {https://doi.org/10.1103/PhysRevLett.127.202301} {\bibfield
  {journal} {\bibinfo  {journal} {Phys. Rev. Lett.}\ }\textbf {\bibinfo
  {volume} {127}},\ \bibinfo {pages} {202301} (\bibinfo {year}
  {2021}{\natexlab{b}})},\ \Eprint {https://arxiv.org/abs/2011.06078}
  {arXiv:2011.06078 [nucl-ex]} \BibitemShut {NoStop}%
\bibitem [{\citenamefont {Aaij}\ \emph {et~al.}(2019)\citenamefont {Aaij} \emph
  {et~al.}}]{LHCb:2018weo}%
  \BibitemOpen
  \bibfield  {author} {\bibinfo {author} {\bibfnamefont {R.}~\bibnamefont
  {Aaij}} \emph {et~al.} (\bibinfo {collaboration} {LHCb}),\ }\bibfield
  {title} {\bibinfo {title} {{Prompt $\Lambda^+_c$ production in p--Pb
  collisions at $\sqrt{s_{NN}} = 5.02$ TeV}},\ }\href
  {https://doi.org/10.1007/JHEP02(2019)102} {\bibfield  {journal} {\bibinfo
  {journal} {JHEP}\ }\textbf {\bibinfo {volume} {02}},\ \bibinfo {pages}
  {102}},\ \Eprint {https://arxiv.org/abs/1809.01404} {arXiv:1809.01404
  [hep-ex]} \BibitemShut {NoStop}%
\bibitem [{\citenamefont {Varga}\ and\ \citenamefont
  {Vertesi}(2022)}]{Varga:2021jzb}%
  \BibitemOpen
  \bibfield  {author} {\bibinfo {author} {\bibfnamefont {Z.}~\bibnamefont
  {Varga}}\ and\ \bibinfo {author} {\bibfnamefont {R.}~\bibnamefont
  {Vertesi}},\ }\bibfield  {title} {\bibinfo {title} {{The role of the
  underlying event in the enhancement in high-energy pp collisions}},\ }\href
  {https://doi.org/10.1088/1361-6471/ac7248} {\bibfield  {journal} {\bibinfo
  {journal} {J. Phys. G}\ }\textbf {\bibinfo {volume} {49}},\ \bibinfo {pages}
  {075005} (\bibinfo {year} {2022})},\ \Eprint
  {https://arxiv.org/abs/2111.00060} {arXiv:2111.00060 [hep-ph]} \BibitemShut
  {NoStop}%
\bibitem [{\citenamefont {Gladilin}(2015)}]{Gladilin:2014tba}%
  \BibitemOpen
  \bibfield  {author} {\bibinfo {author} {\bibfnamefont {L.}~\bibnamefont
  {Gladilin}},\ }\bibfield  {title} {\bibinfo {title} {{Fragmentation fractions
  of $c$ and $b$ quarks into charmed hadrons at LEP}},\ }\href
  {https://doi.org/10.1140/epjc/s10052-014-3250-3} {\bibfield  {journal}
  {\bibinfo  {journal} {Eur. Phys. J. C}\ }\textbf {\bibinfo {volume} {75}},\
  \bibinfo {pages} {19} (\bibinfo {year} {2015})},\ \Eprint
  {https://arxiv.org/abs/1404.3888} {arXiv:1404.3888 [hep-ex]} \BibitemShut
  {NoStop}%
\bibitem [{\citenamefont {Bierlich}\ \emph {et~al.}(2024)\citenamefont
  {Bierlich}, \citenamefont {Gustafson}, \citenamefont {L\"onnblad},\ and\
  \citenamefont {Shah}}]{Bierlich:2023okq}%
  \BibitemOpen
  \bibfield  {author} {\bibinfo {author} {\bibfnamefont {C.}~\bibnamefont
  {Bierlich}}, \bibinfo {author} {\bibfnamefont {G.}~\bibnamefont {Gustafson}},
  \bibinfo {author} {\bibfnamefont {L.}~\bibnamefont {L\"onnblad}},\ and\
  \bibinfo {author} {\bibfnamefont {H.}~\bibnamefont {Shah}},\ }\bibfield
  {title} {\bibinfo {title} {{The dynamic hadronization of charm quarks in
  heavy-ion collisions}},\ }\href
  {https://doi.org/10.1140/epjc/s10052-024-12593-0} {\bibfield  {journal}
  {\bibinfo  {journal} {Eur. Phys. J. C}\ }\textbf {\bibinfo {volume} {84}},\
  \bibinfo {pages} {231} (\bibinfo {year} {2024})},\ \Eprint
  {https://arxiv.org/abs/2309.12452} {arXiv:2309.12452 [hep-ph]} \BibitemShut
  {NoStop}%
\bibitem [{\citenamefont {Altmann}\ and\ \citenamefont
  {Skands}(2024)}]{Altmann:2024odn}%
  \BibitemOpen
  \bibfield  {author} {\bibinfo {author} {\bibfnamefont {J.}~\bibnamefont
  {Altmann}}\ and\ \bibinfo {author} {\bibfnamefont {P.}~\bibnamefont
  {Skands}},\ }\bibfield  {title} {\bibinfo {title} {{String junctions
  revisited}},\ }\href {https://doi.org/10.1007/JHEP07(2024)238} {\bibfield
  {journal} {\bibinfo  {journal} {JHEP}\ }\textbf {\bibinfo {volume} {07}},\
  \bibinfo {pages} {238}},\ \Eprint {https://arxiv.org/abs/2404.12040}
  {arXiv:2404.12040 [hep-ph]} \BibitemShut {NoStop}%
\bibitem [{\citenamefont {Sirunyan}\ \emph {et~al.}(2023)\citenamefont
  {Sirunyan} \emph {et~al.}}]{CMS:2023lgq}%
  \BibitemOpen
  \bibfield  {author} {\bibinfo {author} {\bibfnamefont {A.~M.}\ \bibnamefont
  {Sirunyan}} \emph {et~al.} (\bibinfo {collaboration} {CMS}),\ }\bibfield
  {title} {\bibinfo {title} {{Observation of enhanced long-range elliptic
  anisotropies inside high-multiplicity jets in pp collisions at the LHC}},\
  }\href@noop {} {\bibfield  {journal} {\bibinfo  {journal}
  {CMS-PAS-HIN-21-013}\ } (\bibinfo {year} {2023})}\BibitemShut {NoStop}%
\end{thebibliography}%

\end{document}